# Mobile devices as experimental tools in physics education: some historical and educational background

L. Darmendrail (luis.darmendrail@etu.unige.ch),
A. Gasparini (alice.gasparini@unige.ch),
A. Müller (andreas.mueller@unige.ch)
Physics Education Group, University of Geneva, 13/03/2023

**Abstract**

The present text provides a short, non-technical account of some historical and educational background and, based on this, of the rationale of the use of mobile devices in physics education[1].

## 1. Predecessors and origins of today's mobile communication and information

We begin with a simple problem: how to transfer information from one person to another, with the objective of preserving and carrying data as fast and accurately as possible. We can start – arbitrarily –with papyri (3 000 BCE), an ancient way of sending messages as letters (on a simple sheet), as scroll or even as books. For as long as the object remains undamaged, this form of communication can preserve messages without "corruption". While papyri were still in use, paper was invented in China 100-200 CE using subproducts of wood. This newer means of communication, still hand-written remained for more than a millennium until 1454, when Johannes Gutenberg invented the printing press. Gutenberg's invention constituted a huge leap forward in that it gave many more people access to information such as books printed in series – but the information carrier remained the same: paper.

A revolutionary change was achieved in 1830, when Samuel F. B. Morris invented the telegraph, giving rise to near-instantaneous transfer of information from one city to another. Following the telegraph, the landline telephone appeared in 1876 and would be the main communications device for almost a century, until 1973 when the first mobile phone was developed by Motorola.

In parallel to the information transfer problem, in 1642 the mathematician, physicist, inventor, philosopher, moralist and theologian Blaise Pascal invented what was arguably the first calculator, using a complex system of mechanical assembly. Pascal's invention represents one of the first steps towards computers. In the 1900s, computers based on analogue systems started to arrive; in 1939, these systems evolved to use digital language; the year 1975 saw the appearance of the mobile computer.

At the same time as Pascal's calculator, sensors – i.e. measurement instruments for otherwise inaccessible quantities – started to be developed, including the barometer in 1643. In the 19th century, many others followed, such as magnetometers, gyroscopes, microphones and accelerometers. All these sensors and the evolution of technology led in 1993 to the development of microelectromechanical system (MEMS).

Following the interacting evolution of information, computation and sensors, and if we consider the special ingredient of the internet, developed in the 1980s, we finally arrive at modern smartphones and tablets as mobile devices capable of transferring information worldwide, computing unimaginable numbers of equations and amounts of data, and measuring many quantities like distances, acceleration, sound, pressure, magnetic fields, and even subatomic particles (Keller et al., 2019). Figure 2.1 gives an overview of some of the branches that contributed to the birth of modern mobile devices.

---

[1] Related, among others, to work done at the physics education group the University of Geneva (Darmendrail et al., 2019; Darmendrail & Müller, 2020, 2021a, b, c; Gasparini, 2021; Keller et al., 2019).

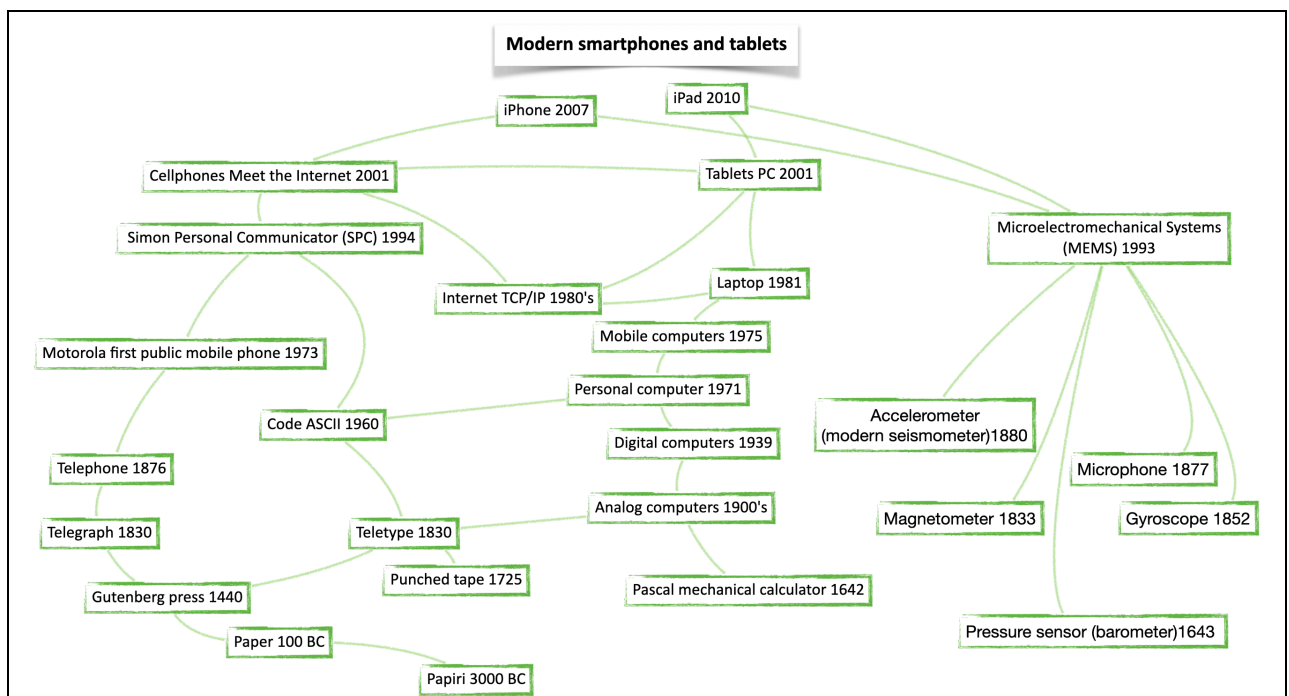

Fig. 2.1: Evolution of different branches of technology that converge in mobile devices.

In the remainder of this text, we will review firstly the application of these mobile technologies in physics (science) education and secondly how they are integrated in physics experiments that drive technology-based learning and its active and interactive perspectives in teaching with mobile devices.

## 2. Science education background

### 2.1. Context-based science education

One typical problem in science education is to establish meaningful contexts for sometimes rather abstract or counterintuitive concepts, especially when these are also connected to a mathematical description, as is often the case in physics. This might give students an impression of physics as a dry, impersonal and irrelevant discipline (Bennett et al., 2007; Kuhn & Müller, 2014), which has nothing to do with their lives nor the world they live in.

Here, context plays a critical role. Basing teaching in everyday life contexts has been a strong strand of R&D in science education for more than three decades (Bennett et al., 2007) and the main objective is to bring closer to reality the so called "dry" theories of physics (science). Context-based science education (CBSE) relates to typical phenomena of everyday life and gives theoretical support for better comprehension. It is in this relation between phenomena and theory that MDETs play an important role, being ubiquitous and able to serve as laboratories in pockets. They are useful not only for processing data but also due to the manifold sensors and applications that can facilitate data collection ad analysis. Activities with MDETs in CBSE can be implemented as a partial or, in some cases (like pandemics), complete alternative of the traditional laboratory, depending on the situation and implementation of the class. The experiments in ch. 4 ("Everyday phenomena measured with mobile devices") and ch. 7 (from an undergraduate course on physics of everyday phenomena) are strongly intended in that sense; further examples can be found in ch. 5.

An example of CBSE is presented by Kuhn & Müller (2014) using newspaper story problems: a trip around the planet in an electric plane serves as an input for questions related to electrical power production and consumption of the plane. This example is one among many others, like sales promotions for home appliances, news about some spectacular flying cars, or claims about a man who jumps from the stratosphere, which can all be verified or rejected using scientific theory.

This implementation of CBSE had a positive impact on students, not only in learning the science content (with medium effect sizes) but also in motivation (with large effect sizes); thus, the "irrelevant" discipline is now closer than expected. These effects are shown in empirical studies like Bennett et al., 2007; Gilbert et al., 2011; Müller et al., 2014.

## 2.2. Hands-on, active, and interactive learning

It is well-known that, in recent decades, the implementation of more practical teaching has taken a prominent place in the teaching of sciences, meaning that it is more fruitful for the acquisition of knowledge to let students work by themselves to produce data by actively constructing experiments and taking measurements with formal instruments or, in the case of this work, with MDETs as example of mobile technologies.

Müller and Brown (2022) write that many practitioners and researchers see hands-on teaching see as a promising means to engage and motivate students. They cite studies by Häussler and colleagues (1998) and Swarat and colleagues (2012) to confirm that interest in science, in particular, is heightened by hands-on activities.

Beyond interest, science achievement is also strongly enhanced by hands-on activity (Wise & Okey, 1983; Schroeder et al., 2007). The value of hands-on activity is not ~~be~~ limited to science, Carbonneau and colleagues (2013) found similar results for "manipulatives" in mathematics education.

The review of Müller and Brown (2022) also shows the importance of hands-on activities as an active way of learning and provide ample evidence in support of this approach. When we want to put the student in a more active position for learning, adding mobile devices as a tool to take data in experiments, can give students a sense of "ownership" (Szott, 2014) because it is by their own measurements (and sometimes devices) that they obtain real, authentic data for further analysis (Song, 2014).

## 2.3. Technology-based learning

Teaching methods underwent many changes throughout the 20th century, as did the use of different media to spark students' interest in science – from a focus on reading books and papers, which still can have a positive impact on students today (Schiefele, 1999; Silvia, 2006), to technology-based learning (Swarat et al., 2012), e.g. using MDETs to foster affective factors, especially towards physics.

Since 2000, the arrival of mobile technologies has profoundly changed the way we interact with other people and how we obtain information. In everyday conversations, we can consult or verify information using a smartphone or tablet connected to the internet. By way of an example of the rapid progress in this area, the number of mobile phone contracts in Switzerland went from 2 per 100 people in 1990 to 127 in 2020 (source: data.worldbank.org).

In this fully connected world, the advantages of using MDETs to teach science lies in the fact that smartphones and tablets have the capacity to simultaneously record and process data. Among others, this alone enables multiple, complementary representations of the same information (Treagust et al., 2017).

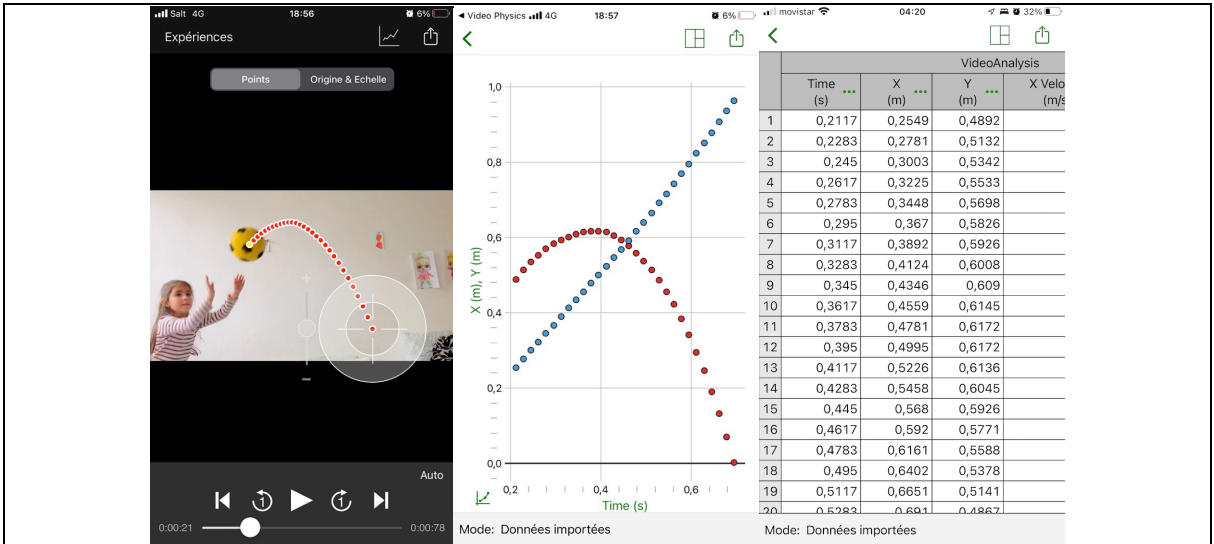

Fig. 2.2: The different representations for the parabolic movement experiment. Form left to right, the actual video of the experiment and the trace of the ball, the graph of position in x and y, and the data table of the experiment

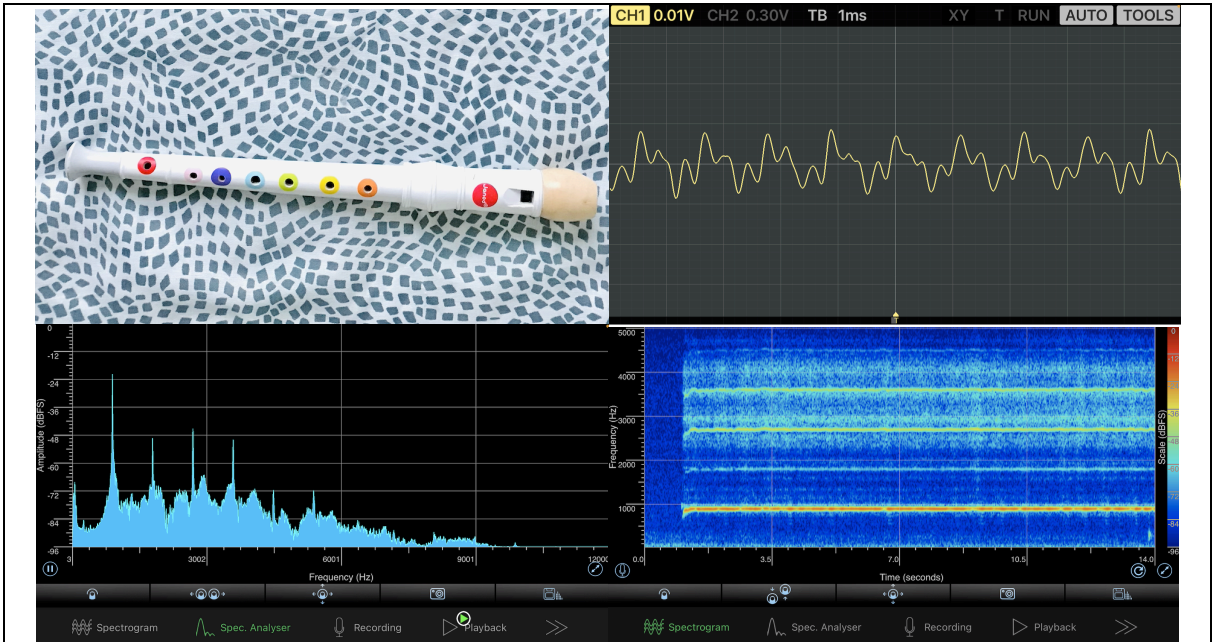

Fig. 2.3: The different representations for a musical sound experiment. Clockwise from top left: the recorder used to produce a sound,; the sound wave of the recorder, the frequency spectrum of the sound, and the spectrum map of the sound.

A first example (see Fig. 2.2) is projectile motion. This two-dimensional movement has a parabolic graph that also matches the real motion of the projectile. The different representations in this case are the video capture in the smartphone, the graph and the tables generated in the application.

A second example (see Fig. 2.3) is music or musical sounds recorded with the microphone of a mobile device. The sound data recorded of an instrument or voice can be represented as a soundwave in an active plot, showing in real time the waveform of the sound, as a frequency spectrum or, in a more complete and complex way, AS a spectrum map where the sound is plotted with information about frequency, amplitude and time in the same time. Finally, this acoustic information can be also represented as numerical data when exporting the record e.g. as a .csv file.

### 2.4. *Teaching with mobile devices*

Considering together that (a) CBSE and (b) active, hands-on learning have been shown a to have positive impact on cognitive and affective outcomes of students, that (c) the use of technology provides further motivational factor, and (d) in view of technology as tool to support cognitive processes (e.g. for multiple representations), a natural conclusion is that MDETs appears as a promising approach in science teaching.

Tablets and smartphones offer to take measurements with new sensors and new applications everywhere. The ability to record and process data in a professional way has been largely restricted to the laboratory, today MDETs offer "portable labs" for many fields of science, especially in subdomains of physics such as radioactivity, electricity, acoustics, optics and mechanics, among others.

The classic demonstration of a ball rolling down an inclined plane traditionally required photo cells, data sheets and calculators or PCs. Today, with a smartphone and the right application, this classical experiment can be performed using video recording and analysis in only a few minutes. In addition, multiple representations of the data can be produced simultaneously (video, graph, table), giving students more time to process the information and discuss more efficiently about their questions about the experiment and the different representations at hand.

Many experiments have been suggested in this vein. The journal *The Physics Teacher* maintains a section devoted to this very subject since 2012 (Kuhn & Vogt, 2012). Many examples have been presented using smartphones and tablets as the main tool to obtain and process data in experiments in science (Darmendrail et al., 2019) showing the potential of MDETs in the teaching of physics in many fields, including:

- mechanics (Chevrier et al., 2013; Gabriel & Backhaus, 2013; Darmendrail & Müller, 2021; Hochberg et al. 2016; Koleza & Pappas, 2008; Kuhn & Vogt, 2013; Monteiro et al., 2014; Shakur & Sinatra, 2013)
- waves and acoustics (Castro-Palacio & Velázquez-Abad, 2013; Parolin & Pezzi, 2013; Sans et al., 2013; Forinash & Wisman, 2012; Greenslade, 2016; Hirth et al., 2016; Vogt et al., 2015; Darmendrail & Müller, 2020; Darmendrail & Müller, 2021)
- optics (Klein, et al., 2014; Thoms, et al., 2013)
- electromagnetism and radioactivity (Forinash & Wisman, 2012; Silva, 2012; Kuhn et al., 2014; Keller et al., 2019).

From a teacher's perspective, the use of MDETs considerably enhances their freedom in laboratory work, enlarging the range of possible student experiments, and sometimes requiring less budget, effort, preparation and time as conventional experiments. Note that it is not the claim that MDETs can offer a one-and-for-all solution for all classroom-experimentation: they also need maintenance and replacements, and there are certainly cases where the traditional material is superior. On the other hand, it is worth pointing out that MDETs (i) complete the choice of tools to teach physics (the sciences) in a contextualized way (i) allow a new approach to teaching, which is halfway between the classic laboratory and exercise sessions (faster than a traditional lab, more authentic than just paper and penci excercise) and (ii) the MDETs

In conclusion, there is a considerable potential of MDETS on the educational level: (a) CBSE and (b) active, hands-on learning have been shown a to have positive impact on cognitive and affective outcomes of students; (c) the use of technology provides a further motivational factor, as well as (d) the sense of ownership of data (i.e. experiment, data collection and analysis is more directly in the hands of learner themselves); (d) MDETs provide a technological tool to support cognitive processes (e.g. for multiple representations).